\documentclass[final,onecolumn]{aipproc}
\layoutstyle{8x11single}

\usepackage{graphicx}
\usepackage{amsmath}
\usepackage{amssymb}
\usepackage{caption}

\newcommand{\trac}{{\rm Tr}}
\newcommand{\Tr}{{\rm Tr}}
\newcommand{\diag}{{\rm diag}}

\newcommand{\Boltz}{ k_{\rm\scriptscriptstyle B}}

\newcommand{\be}{\begin{equation}}
\newcommand{\ee}{\end{equation}}
\newcommand{\bea}{\begin{eqnarray}}
\newcommand{\eea}{\end{eqnarray}}
\newcommand{\lt}{\left(}
\newcommand{\rt}{\right)}
\newcommand{\lgr}{\left\{}
\newcommand{\rgr}{\right\}}
\newcommand{\fr}{\frac}

\newcommand{\onehalf}{{\mbox{\textonehalf\,}}}
\newcommand{\quot}[1]{{``#1''}}

\begin{document}

%%%%%  The title of your contribution in all capital letters.
\title{STEEPEST ENTROPY ASCENT PATHS\\ TOWARDS THE MAX-ENT DISTRIBUTION}

%%%%%  Author's name (underline presenter's author)
\author{Gian Paolo Beretta}{address={
        Universit\`a di Brescia, via Branze 38, 25123
Brescia, Italy}, , email={beretta@ing.unibs.it}
       }


\date{\today}

%%%%%               Abstract begins here
\begin{abstract}
With reference to two general probabilistic description frameworks,  Information Theory and  Classical Statistical Mechanics, we discuss the geometrical reasoning and mathematical formalism leading to the differential equation that defines  in probability space the smooth path of Steepest Entropy Ascent (SEA)  connecting an arbitrary initial probability distribution to the unique Maximum Entropy (MaxEnt) distribution with the same mean values of a set of  constraints. The SEA path is relative to a metric chosen to measure distances in square-root probability distribution space. Along the resulting SAE path, the metric turns out to be proportional to the concept of Onsager resistivity generalized to the far  non-equilibrium domain. The length of the SEA path to MaxEnt provides a novel global measure of degree of disequilibrium (DoD) of the initial probability distribution, whereas a local measure of DoD is given by the norm of a novel generalized concept of non-equilibrium affinity.
\end{abstract}

\maketitle

\section{Introduction}

Various fields  often benefited from tools and concepts originally developed within thermodynamics.  In the hope that it finds applications also in other fields, in this paper we present the essential  features of a formulation that allows to describe and compute the smooth change of an initially arbitrary probability distribution as it evolves towards MaxEnt along one a path which preserves the mean values of a set of constraints. Among the innumerable such paths, our construction selects and defines the smooth path of Steepest Entropy Ascent (SEA) with respect to a prescribed proper metric field in the space of probability distributions.
   In the non-equilibrium thermodynamics framework,  standard near-equilibrium results of the linear theory of irreversible processes obtain \cite{JETC2013} when the  metric field at MaxEnt is proportional to  matrix of generalized Onsager resistivities.

In the first two Sections we introduce slightly nonstandard notations which allow the unified  SEA construction and implementation we present in the third Section  for the two non-equilibrium frameworks we focus on in this paper. The final Section discusses the main results including natural definitions of local and global measures of degree of disequilibrium for an arbitrary probability distribution.

\section{\label{A}Framework A: Information Theory}

Let $ \mathcal{L} $ be the set of all $ n \times n $ real,
diagonal matrixes $ A = \diag ( a_j )$, $B = \diag ( b_j )$, \dots
( $n \leq \infty$ ), equipped with the inner product $ ( \cdot |
\cdot ) $ defined by \be \lt A | B \rt = \trac (A B) = {\textstyle
\sum_{j = 1}^n} a_j\, b_j \label{1c} \ee \noindent
In Information Theory \cite{Jaynes}, the probability assignment to a set of $ n $ events, $
p_j $ being the probability of occurrence of the $ j $-th event, is represented by $ \rho = \diag ( p_j ) $. Again, in order to easily impose the constraint of preservation of the nonnegativity of the probabilities during their time evolution, we adopt the description in terms of the square-root of $\rho$ that we denote by
\be \gamma = \diag ( \sqrt{p_j} ) \label{2c}\ee

Typically we consider a set of conserved features of the process \be\lgr C_i \rgr= \lgr H , N_1 ,
\ldots , N_r, I \rgr    \ee  of diagonal matrixes $ H = \diag ( e_j )$,
$N_1 = \diag ( n_{1 j } )$, \dots, $N_r = \diag ( n_{rj} ) $, $I = \diag ( 1 ) $ in $
\mathcal{L} $ representing  characteristic features of the
events in the set, which for the $ j $-th event take on respectively the values $ e_j $, $n_{1 j} $, \dots , $n_{r j} $. The corresponding expectation values are $\trac (\rho H)={\textstyle \sum_{j = 1}^n} p_j\,e_j $, $\trac (\rho N_1)={\textstyle \sum_{j = 1}^n} p_j\,n_{1 j}$, \dots, $\trac (\rho N_r)={\textstyle \sum_{j = 1}^n} p_j\,n_{r j}$, and $\trac (\rho I)={\textstyle \sum_{j = 1}^n} p_j = 1$ thus providing the normalization condition on $\rho$.

The time evolution of the square-root probability distribution  $\gamma$ is the solution of the rate equation
\be \frac{d\gamma}{dt} = \Pi_{\gamma} \label{3c} \ee
where in order to satisfy the constraints of conservation of the expectation values $\trac (\rho C_i) =  \onehalf (\Psi_i |\gamma)$ the  term $\Pi_{\gamma}$ must be such that
\be \Pi_{C_i}=\frac{d}{dt}\trac(\rho C_i)=  (\Psi_i |\Pi_{\gamma})  = 0  \qquad {\rm where} \qquad  \Psi_i=2\gamma C_i \label{4c} \ee

The entropy functional  in this context is represented by
\be S ( \gamma ) = - \Boltz \trac (\rho \ln \rho) =  (-\Boltz\gamma\ln\gamma^2 |\gamma) = \onehalf (\Phi |\gamma) \label{5c} \ee
so that the rate of entropy production under a time evolution that preserves the normalization of $\rho$ is given by
\be \Pi_{S}=- \Boltz \frac{d}{dt}\trac(\rho \ln \rho)=  (\Phi |\Pi_{\gamma}) \qquad {\rm where} \qquad \Phi=-2\Boltz\gamma\ln \gamma^2 \label{6c} \ee
Below,  we  construct a model for the rate term $\Pi_{\gamma}$ such that $\Pi_S$ is maximal subject to the conservation constraints $\Pi_{C_i}=0$ and a suitable additional constraint.
An attempt along similar lines has been presented in \cite{Lemanska}.

\section{\label{B}Framework B: Classical Statistical Mechanics}

Let $ \Omega $ be the classical position-momentum $q$--$p$ phase space, and $ \mathcal{L} $ the set of
real, square-integrable functions $ A , B , \ldots $ on $ \Omega
$, equipped with the inner product $ ( \cdot | \cdot ) $ defined
by \be (A | B) = \trac (A B) = {\textstyle \int_{\Omega}}\; A B \,\,
d \Omega \label{1a} \ee \noindent where $ \trac (\cdot) $ in this
framework denotes $ \int_{\Omega} \cdot\,d \Omega $, with $d \Omega =d\textbf{q}\,d\textbf{p}$.

%We denote by
%$ \mathcal{P} $ the subset of all nonnegative-definite
%functions  $ \rho $ in $ \mathcal{L} $, i.e., \be
%\mathcal{P} = \big\{ \rho \mbox{ in } \mathcal{L}\, |\, \rho \geq
%0 \big\} \label{2a} \ee \noindent

In Classical Statistical Mechanics, the index
of statistics from a generally heterogeneous ensemble of identical
systems (with associated phase space $ \Omega $) distributed over
a range of possible classical mechanical states is represented by a  nonnegative (Gibbs)
density-of-phase distribution function $ f_{\rm G}=f_{\rm G}(\textbf{q},\textbf{p},t) $ in $ \mathcal{L} $.

Borrowing from the formalism we originally developed for the quantum framework \cite{LectureNotes}, in order to easily impose the constraint of preservation of the nonnegativity of $f_{\rm G}$ during its time evolution, we adopt as state representative not $f_{\rm G}$ itself but its square root, that we assume to be a function in $ \mathcal{L} $ that denote by $ \gamma=\gamma(\textbf{q},\textbf{p},t) $. Normalization is not imposed at this stage but later as one of the constraints.  Therefore, we clearly have
\be f_{\rm G} = \gamma^2 \ ,\qquad \frac{\partial f_{\rm G}}{\partial t}=2\gamma\frac{\partial\gamma}{\partial t}\ ,\qquad  \frac{\partial f_{\rm G}}{\partial \textbf{q}}=2\gamma\frac{\partial\gamma}{\partial \textbf{q}}\ ,\quad \frac{\partial f_{\rm G}}{\partial \textbf{q}}=2\gamma\frac{\partial\gamma}{\partial \textbf{q}}\ ,\quad  \{H,f_{\rm G}\} = 2\gamma \{H,\gamma\} \ee where
$\{\cdot,\cdot\}$ denotes the Poisson bracket.

Among the phase-space functions representing physical observables we focus on the conserved ones that we denote synthetically by the set \be\lgr C_i \rgr =
\lgr H , M_x , M_y , M_z, N_1 , \ldots , N_r, I \rgr \ee
where $ H $ is the classical Hamiltonian function, $M_j$ the momentum function for the $j$-th component, $ N_i $ the
number-of-particle function for particles of type $ i $, and $I=1$ is the constant unity function, so that corresponding mean values are $\trac(\gamma^2 H)$, $\trac(\gamma^2 \textbf{M})$,  $\trac(\gamma^2 N_i)$, and $\trac(\gamma^2 I)$  (normalization).

The description of an irreversible diffusion-relaxation process in this framework can be done by assuming a evolution equation for the state $f_{\rm G}$ given by
\be \frac{d\gamma}{dt} = \Pi_{\gamma} \quad {\rm where}\quad \frac{d}{dt} = \frac{\partial}{\partial t} -\{H,\cdot\} \label{3a} \ee
It is easy to verify that for $\Pi_{\gamma}=0$ Eq.\ (\ref{3a}) reduces to Liouville's equation of classical reversible evolution. We do not make this assumption because we are interested in modeling irreversible evolution with energy, momentum, and particle numbers redistribution towards equilibrium, subject to the overall conservation of energy, momentum, number of particles of each kind, and normalization (notice that also here $\trac(\gamma^2 I)=\onehalf (\Psi_i |\gamma)$)
\be \Pi_{C_i}=\frac{d}{dt}\trac(\gamma^2 C_i)=  (\Psi_i |\Pi_{\gamma}) = 0 \qquad {\rm where} \qquad \Psi_i= 2\gamma C_i \label{4a} \ee

The entropy state functional in this context is represented by
\be S ( \gamma ) = - \Boltz \trac (f_{\rm G} \ln f_{\rm G}) = (-\Boltz\gamma\ln\gamma^2 |\gamma)=\onehalf (\Phi |\gamma)  \label{5a} \ee
so that the rate of entropy production under a time evolution that preserves the normalization of $f_{\rm G}$ is given by
\be \Pi_{S}=- \Boltz \frac{d}{dt}\trac(f_{\rm G} \ln f_{\rm G})=  ( \Phi|\Pi_{\gamma})   \qquad {\rm where} \qquad \Phi= -2\Boltz\gamma\ln \gamma^2 \label{6a} \ee

In  the next Section we  construct an equation of motion for the square-root-of-density-of-phase distribution $\gamma$ such that $\Pi_S$ is maximal subject to the conservation constraints $\Pi_{C_i}=0$ and a suitable additional constraint.

\section{\label{SEA}Steepest Entropy Ascent geometrical construction}

By reformulating the dynamical description in terms of the square-root probabilities $\gamma$, we need not impose the constraint of preservation of  nonnegativity of $\gamma^2$ during the time evolution.
Moreover, the state representation in terms of an element $\gamma$ of a suitable vector space $ \mathcal{L} $ equipped with an inner product $ (\cdot|\cdot)$ allows a unified formulation of frameworks A and B. The term in the dynamical equation for $\gamma$ which is responsible for  dissipative irreversible relaxation and hence entropy generation is the vector $|\Pi_\gamma)$ of $ \mathcal{L} $. Geometrically, in framework A the vector $|\Pi_\gamma)$ represents the vector tangent to the path $\gamma(t)$, and in framework B the dissipative component of the vector tangent to the path $\gamma(t)$.

In both frameworks, the rate of entropy production is given by the scalar product of $|\Pi_\gamma)$ with the vector $|\Phi)$ which represents the  entropy gradient in state space, as it is the variational derivative of the entropy functional with respect to the state vector $\gamma$,
 \be \Pi_S=(\Phi|\Pi_\gamma) \label{1g} \ee
Similarly, the rates of production of each conserved property $C_i$ is given by the scalar product of $|\Pi_\gamma)$ with the vector $|\Psi_i)$ which represents the  gradient of $C_i$  in state space and is the variational derivative of the $C_i$ mean functional with respect to  $\gamma$,
  \be \Pi_{C_i}=(\Psi_i|\Pi_\gamma) \label{2g} \ee

 Our objective is to construct a smooth path, i.e., a continuous one-parameter family $\gamma(t)$, such that $\gamma(0)$ is the given arbitrary initial state and $\gamma(\infty)$ the MaxEnt state such that $(C_i|\gamma(t))=(C_i|\gamma(0))$ for every $t$ from 0 to $\infty$ where the phase-space functions $C_i$ represent the conservation constraints including normalization.

The SEA path we seek is identified by the solution of the following variational problem: find the instantaneous \quot{direction} of    $|\Pi_\gamma)$ which maximizes the entropy production rate $\Pi_S$ subject to the constraints $\Pi_{C_i}=0$. Clearly, we are not interested in the unconditional maximization of  $\Pi_S$ that we could obtain trivially by simply increasing the \quot{norm} of $|\Pi_\gamma)$ indefinitely. Therefore, we must impose the additional constraint that the norm of $|\Pi_\gamma)$ be fixed, $\|\Pi_\gamma\|=\rm prescribed$. Indeed, since $\|\Pi_\gamma\|\,{\rm d}t$ approximates the arc-length  ${\rm d}\ell$ of the path   traveled by $\gamma$  from $\gamma(t)$ to $\gamma(t+{\rm d}t)$, the additional constraint, $\|\Pi_\gamma\|={\rm d}\ell/{\rm d}t=\rm prescribed$, means that among all the conceivable paths that go through $\gamma(t)$, are compatible with the constraints $\Pi_{C_i}=0$, and travel a given small prescribed distance ${\rm d}\ell$ in the time ${\rm d}t$, we select the one that in the same lapse of time  ${\rm d}t$ produces the highest entropy increase, $\Pi_S\,{\rm d}t$.

This important constraint cannot be formalized unless we specify  a metric with respect to which to measure distances between square root probability distributions over the manifold defined by the constraints.
The most general expression for the distance traveled along a path on such a manifold is \be\label{Gell} {\rm d}\ell = 2\,\sqrt{(\Pi_\gamma|\,\hat G\,|\Pi_\gamma)}\, {\rm d}t \ee  where $\hat G$ is a real, symmetric, and positive-definite metric tensor.

Once $\hat G$ is identified, the variational problem is solved by recalling Eqs. (\ref{1g}) and  (\ref{2g}) and introducing Lagrange multipliers $\Boltz\beta_i$ and $\Boltz\tau/4$ to seek the unconstrained maximum, with respect to $\Pi_\gamma$, of the Lagrangian
     \be  \Upsilon= \Pi_S - \Boltz\sum_i \beta_i\, \Pi_{C_i} - \Boltz\tau\, (\Pi_\gamma|\,\hat G\,|\Pi_\gamma) \label{8g1} \ee
The Lagrange multipliers  must be independent of $\Pi_\gamma$ but can be functions of the state $\gamma$. Taking the variational derivative of $\Upsilon$ with respect to $|\Pi_\gamma)$ and setting it equal to zero we finally obtain the SEA  evolution equation
  \be |\Pi_\gamma)=\frac{1}{\Boltz\tau}\hat G^{-1}\,|\Phi -\Boltz\sum_j \beta_j\, \Psi_j)\label{10g} \ee

   Inserting Eq.\ (\ref{10g}) into the conservation constraints (\ref{2g}) yields the important system of equations which defines the values of the Lagrange multipliers $\beta_j$,
  \be  \Boltz\sum_j  (\Psi_i|\,\hat G^{-1}\,|\Psi_j)\,\beta_j =(\Psi_i|\,\hat G^{-1}\,|\Phi)\label{11g} \ee
This system can be readily solved for the $\beta_j$'s (for example by Cramer's rule)  because the  functionals $(\Psi_i|\hat G^{-1}|\Psi_j)$ and $(\Psi_i|\hat G^{-1}|\Phi)$ are readily computable for the current state $ \gamma$. When Cramer's rule is worked out explicitly, the SEA equation (\ref{10g}) takes the form of a ratio of determinants with which we  presented it in the Quantum Thermodynamics framework \cite{thesis,Cimento1,LectureNotes,ROMP,Entropy2008} (except that there we adopted the uniform, Fisher-Rao metric $\hat G=\hat I$)
$$\label{S'gram}|\Pi_\gamma)= \frac{1}{\Boltz\tau}\fr{\left|
\begin{array}{cccc} \hat G^{-1}\,|\Phi) & \hat G^{-1}\,|\Psi_1) & \cdots & \hat G^{-1}\,|\Psi_c) \\ (\Psi_1|\,\hat G^{-1}\,|\Phi)
 & (\Psi_1|\,\hat G^{-1}\,|\Psi_1) &\cdots &  (\Psi_1|\,\hat G^{-1}\,|\Psi_c)
\\
\vdots &\vdots &\ddots &\vdots
\\
 (\Psi_c|\,\hat G^{-1}\,|\Phi) &
(\Psi_c|\,\hat G^{-1}\,|\Psi_1)  &\cdots & (\Psi_c|\,\hat G^{-1}\,|\Psi_c)
\end{array} \right|}{\left|
\begin{array}{ccc} (\Psi_1|\,\hat G^{-1}\,|\Psi_1) &\cdots &  (\Psi_1|\,\hat G^{-1}\,|\Psi_c) \\
\vdots  &\ddots &\vdots
\\
(\Psi_c|\,\hat G^{-1}\,|\Psi_1) &\cdots &  (\Psi_c|\,\hat G^{-1}\,|\Psi_c) \end{array} \right|}
$$

A geometrical illustration of the above construction  is given in Figure 1.

   \begin{figure}[t]
 %\resizebox{.7\columnwidth}{!}
      \centering
        {\includegraphics[width=0.5\textwidth]{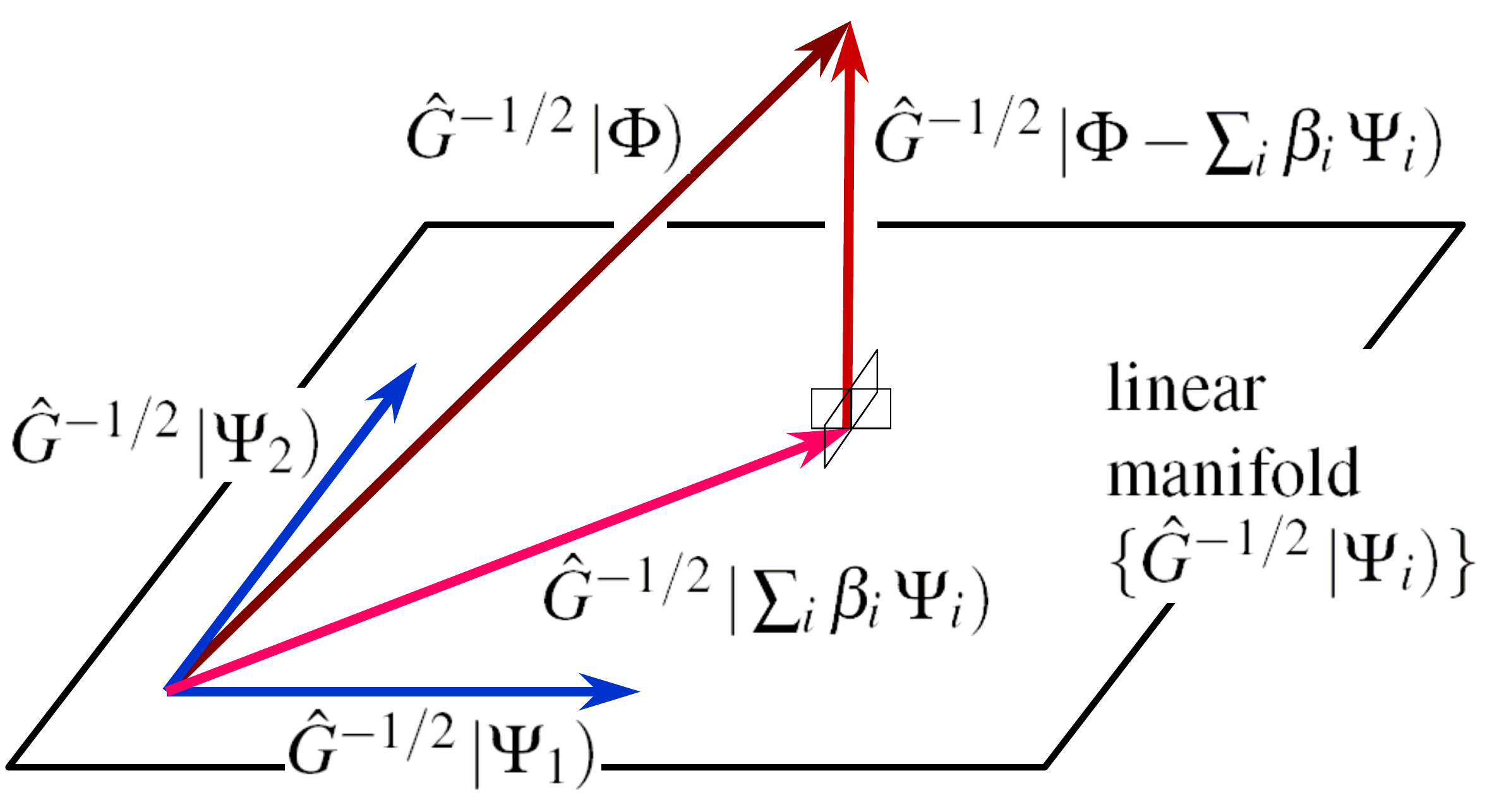}}
        \captionsetup{width=0.5\textwidth}
   \caption{\label{Figure5}\small Pictorial representation of the linear manifold spanned by the vectors $\hat G^{-1/2}\,\Psi_i$ and the orthogonal projection of $\hat G^{-1/2}\,|\Phi)$ onto this manifold which defines the Lagrange multipliers $\beta_i$ for a given  metric tensor $\hat G$. For simplicity, the illustration assumes $\Boltz=1$. The construction defines also the generalized affinity vector $\hat G^{-1/2}\,| \Phi-\sum_i \beta_i\, \Psi_i)$.}
   \end{figure}

 \section{\label{G}Generalized non-equilibrium  affinity, degree of disequilibrium, and final remarks}

In general, the vector
\be |\Lambda)= |\Phi-\Boltz\sum_i \beta_i\, \Psi_i)\label{18g}\ee
may be interpreted as the vector of \quot{generalized  affinities}.  The variational problem and its solution, essentially state that the \quot{rotated tangent vector} $\hat G^{1/2}\, |\Pi_\gamma)$ must have the same direction as that of the \quot{counter-rotated} generalized affinity vector, $\hat G^{-1/2}\, |\Lambda)$ as illustrated in Figure 2. In terms of this vector, Eq.\ (\ref{10g}) rewrites as
 \be |\Pi_\gamma)=\hat L\,|\Lambda) \qquad \mbox{where we define} \qquad \hat L=\frac{1}{\Boltz\tau}\hat G^{-1} \label{19} \ee
 Tensor $\hat L$ represents the field (over state space) of \quot{Onsager generalized conductivities} and its symmetry is a trivial consequence of the symmetry of any proper metric tensor. Hence, we see that the SEA construction entails an Onsager's reciprocity theorem generalized to the entire non-equilibrium domain (not only near equilibrium).

%
%
% Indeed, we first notice that $|\Phi)=2\delta S(\gamma)/\delta|\gamma)$ and $|\Psi_i)=2\delta C_i(\gamma)/\delta|\gamma)$. Moreover, the $\beta$'s are so chosen as to
%
% \be \Pi_S=(\Phi|\Pi_\gamma) = (\Phi-\Boltz\sum_j \beta_j\, \Psi_j |\Pi_\gamma) =2(\Lambda |\,\hat L\,|\Lambda)\ge 0 \label{12g}\ee
%
% \be \delta \Pi_S/\delta |\Pi_\gamma)=2L\,|\Lambda)\ee
%
% \be <A>=\trac(Ag^2) \quad <\dot A>=2\trac(A\dot g g) \quad \delta <A>/\delta g = 2gA \quad <\dot A>=
% \delta <A>/\delta g \cdot \dot g \quad \dot g = \ee

 \begin{figure}[t]
     % \resizebox{.7\columnwidth}{!}
      \centering
       {\includegraphics[width=0.5\textwidth]{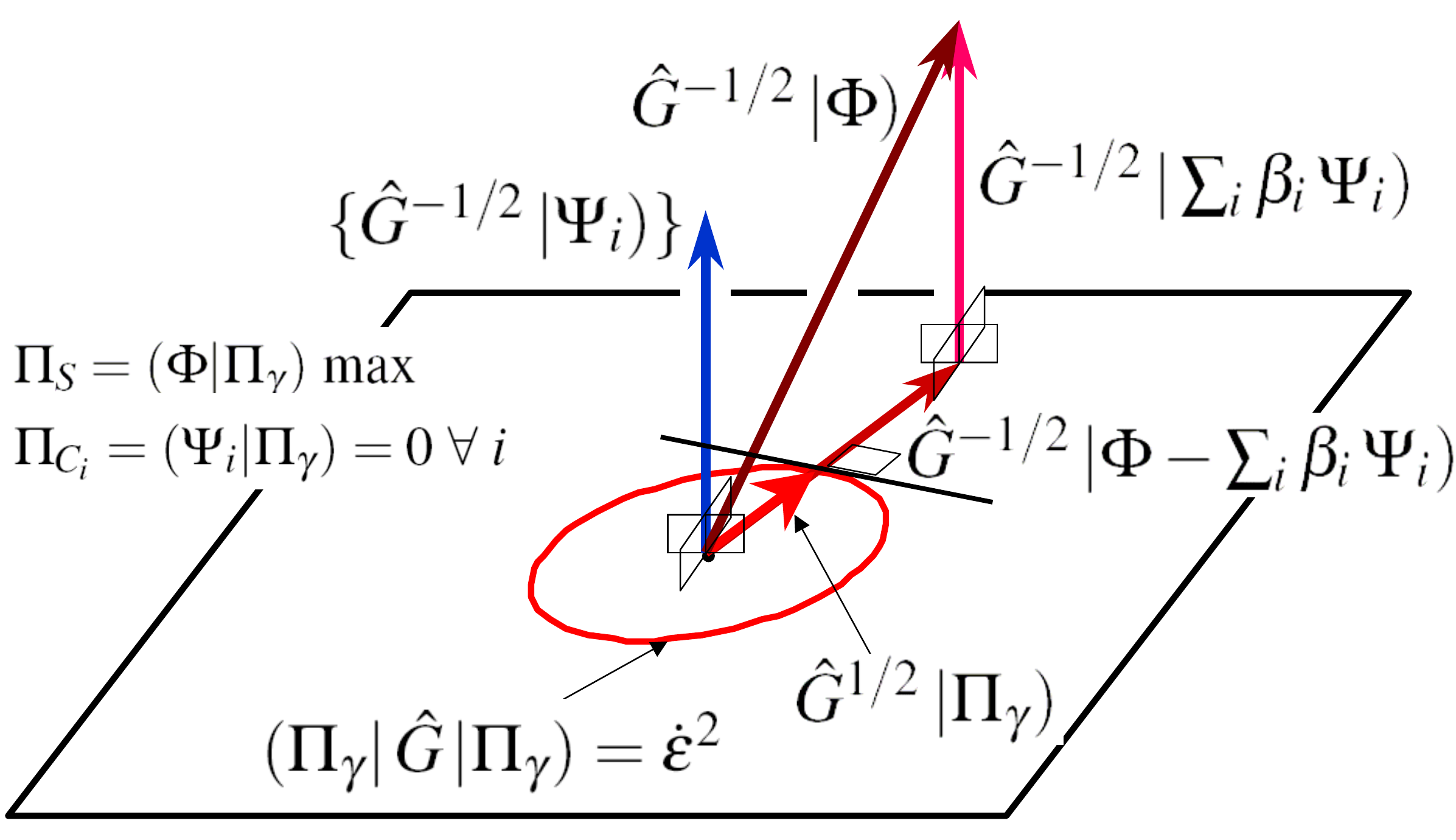}}
       \captionsetup{width=0.5\textwidth}
 \caption{ \label{Figure3}\small Pictorial representation of the SEA variational construction for a given metric tensor $\hat G$. The circle  represents  the  condition $(\Pi_\gamma|\,\hat G\,|\Pi_\gamma)=\rm const$, corresponding to the norm of vector $\hat G^{1/2}\,|\Pi_\gamma)$. This vector must be orthogonal to the $\hat G^{-1/2}\,|\Psi_i)$'s in order to satisfy the conservation constraints $\Pi_{C_i}=(\Psi_i|\Pi_\gamma)=0$. In order to  maximize the scalar product  $\Pi_{S}=(\Phi|\Pi_\gamma)=(\Phi-\sum_i \beta_i\, \Psi_i|\Pi_\gamma)$, vector $\hat G^{1/2}\,|\Pi_\gamma)$ must have the same direction as $\hat G^{-1/2}\,| \Phi-\sum_i \beta_i\, \Psi_i)$. Also here, for simplicity, the illustration assumes $\Boltz=1$.}
   \end{figure}

The thermodynamic principle of non-negative entropy production ($H$-theorem) is easily proved. Subtracting  Eqs.\ (\ref{2g}), each multiplied by the corresponding $\beta_j$,  from Eq.\ (\ref{1g}) and then inserting Eq.\ (\ref{10g}) yields the following explicit expressions for  rate of entropy production
  \be \Pi_S=(\Phi|\Pi_\gamma) = (\Phi-\Boltz\sum_j \beta_j\, \Psi_j |\Pi_\gamma) =(\Lambda |\,\hat L\,|\Lambda)\ge 0 \label{12g}\ee
which is clearly nonnegative-definite since so is $\hat L$.

When only some of the components (partial affinities) of  vector $\Lambda$ are zero, the state is partially equilibrated (equilibrated with respect to the corresponding underlying components of the state $\gamma$). When $\Lambda=0$, then and only then we have an equilibrium state or a non-dissipative ( limit cycle. In fact, that is when and only when the entropy production vanishes.

The norm of the vector  $|\Phi -\sum_j \beta_j\, \Psi_j)$ with respect to the inverse of the  metric tensor $\hat G$, can be adopted as a local measure of the \quot{overall degree of disequilibrium}, that we may denote as DoD, \be {\rm DoD}= (\Lambda|\,\hat G^{-1}\,|\Lambda) \ee  Clearly, this definition is valid no matter how far the state is from the (maximum entropy) stable equilibrium state, i.e., also for highly non-equilibrium states. Moreover, it makes  the entropy production rate $\Pi_S$ to appear just proportional to the DoD via the intrinsic overall relaxation time $\tau$, \be \Pi_S=\frac{\rm DoD}{\Boltz\tau} \ee Of course, also $(\Lambda|\Lambda)$, the norm  of $|\Lambda)$ with respect to the uniform metric,  can be viewed as another measure of the \quot{degree of disequilibrium}, as well as any  norm with respect to other metrics.

Another natural measure of the degree of disequilibrium is the following global
 definition of \quot{distance from MaxEnt along the SAE path} provided by the SEA construction. For any distribution $\gamma$,  the overall length of the SEA path from $\gamma(0)=\gamma$ to $\gamma(\infty)=\gamma_{{\rm MaxEnt}_{C_i(\gamma)}}$
  \be d_{\rm SEA}(\gamma,\gamma_{{\rm MaxEnt}_{C_i(\gamma)}}) = 2\int_0^\infty \sqrt{(\Pi_\gamma|\,\hat G\,|\Pi_\gamma)}\, dt \ee
  Of course, such definition is relative to the particular metric tensor field $\hat G$ which may in general be a function of the state $\gamma$. For the uniform (Fisher-Rao) metric tensor $\hat G=\hat I$ it defines the length of the geodesic connecting $\gamma$ to $\gamma_{{\rm MaxEnt}_{C_i(\gamma)}}$.
Except near MaxEnt, this is of course different than the degree of disequilibrium measured by the Kullback–Leibler divergence or relative entropy of the probability distribution with respect to the MaxEnt distribution with the same mean values of the constraints, which in our notation is
\be \Tr[ \gamma^2 \ln (\gamma^2 /\gamma^2_{{\rm MaxEnt}_{C_i(\gamma)}}) ]\ee

The   speed $d\ell/dt$ at which the state $\gamma$ evolves along the SEA path (which we held fixed in our construction aimed at identifying the SEA direction) is obtained by inserting Eq.\ (\ref{10g}) into  (\ref{Gell}),
\be\frac{d\ell^2}{dt^2}=4\,(\Pi_\gamma|\,\hat G\,|\Pi_\gamma)=
  \frac{(\Lambda |\,\hat G^{-1}\,|\Lambda)}{\Boltz^2\tau^2}=
  \frac{(\Lambda |\,\hat L\,|\Lambda)}{\Boltz\tau}=\frac{\Pi_S}{\Boltz\tau} \label{13g} \ee
 from which we see that the Lagrange multiplier $\tau$ is related to the entropy production rate and the speed $d\ell/dt$. In other words, after the metric $\hat G$ is specified,  through $\tau$ we may specify either the speed at which $\gamma$ evolves along the SEA trajectory in state space or the instantaneous rate of entropy production. Conversely, rewriting Eq.\ (\ref{13g}) as
 \be \Boltz\tau=\frac{\sqrt{(\Lambda |\,\hat G^{-1}\,|\Lambda)}}{d\ell/dt}=\frac{(\Lambda |\,\hat G^{-1}\,|\Lambda)}{\Pi_S}\label{14g} \ee
 The relaxation time $\tau$  needs not be constant. It may be a state functional. For example,
 we see that using $\tau(\gamma)$ as given by Eq.\ (\ref{14g}) the evolution equation  Eq.\ (\ref{10g}) will produce a SEA trajectory in state space with a prescribed entropy production $\Pi_S(\gamma)$. Eq.\ (\ref{14g}) also clearly supports the interpretation of $\tau$ as the \quot{overall relaxation time}.  In any case, even if $\tau$ is constant, the SEA principle provides a  nontrivial non-linear evolution equation that is well defined and reasonable even far from equilibrium.

 We finally note that in the two frameworks we consider here, the only contribution to the entropy change comes from the production term $\Pi_S$, i.e.,  the entropy balance equation  is  $dS/dt=\Pi_S$. Then, Eq.\ (\ref{13g} ) may be rewritten as
  \be \frac{d\ell}{dt/\tau}=\frac{dS}{d\ell}\label{15g} \ee
  from which we see that when time $t$ is measured in units of the \quot{intrinsic relaxation time} $\tau$ the "speed" along the SEA trajectory  is equal to the local entropy gradient  along the trajectory.

The SEA construction presented here is a generalization of a similar construction proposed by the author for  the quantum thermodynamics framework \cite{LectureNotes,ROMP,Entropy2008}, in which so far we adopted the uniform (Fisher-Rao) metric. Recently, however, we noted \cite{JETC2013} that SEA dynamics with respect to a more general metric $\hat G$ allows to capture also the essential elements of the framework of Mesoscopic Non-Equilibrium Thermodynamics \cite{Mazur,BedeauxMazur} as well as of several other non-equilibrium modeling frameworks. Standard results such as the Fokker-Planck equation and Onsager theory emerge as straightforward near-equilibrium results  of SEA dynamics, when the metric  tensor near the maximum entropy state is taken to be directly proportional to the Onsager generalized resistivity tensor, which in turn represents, in the near-equilibrium regime, the strength of the system's reaction when pulled away from equilibrium, i.e., the strength of its tendency to return to MaxEnt. We therefore hope that the SEA construction may turn out to be helpful to  generalize  such standard near-equilibrium results to highly non-equilibrium states in a thermodynamically consistent way, i.e., consistent with the principle of non-negative entropy production.

%  \section{Varie}
%
%  $$\langle A \rangle=(\gamma |A|\gamma)$$
%    $$\frac{d\langle A \rangle}{dt}=(\Pi_\gamma |A|\gamma)+$$
%$$|\Pi_\gamma)=L |\Phi)$$

\bibliographystyle{unsrt}

%%%%%        END of Key Words

\thispagestyle{empty}
\end{document}